\documentclass[
reprint,
amsmath,amssymb,
aps,
prb,
floatfix,
]{revtex4-2}

\usepackage{graphicx}
\usepackage{dcolumn}
\usepackage{bm}
\usepackage{tikz}
\usepackage{ bbold }

\newcommand{\Ha}{\mathcal{H}}
\newcommand{\psid}{\hat{\psi}^\dagger}

\begin{document}

\preprint{APS/123-QED}

\title{Entangled magnon-pair generation in a driven synthetic antiferromagnet}

\author{A.L. Bassant$^1$}
 \email{a.l.bassant@uu.nl.}
 
\author{M.E.Y. Regout$^2$}
\author{J.S. Harms$^1$}
\author{R.A. Duine$^{1,3}$}
\affiliation{$^1$Institute for Theoretical Physics, Utrecht University, Princetonplein 5, 3584CC Utrecht, The Netherlands \\
$^2$Nanomat, Université de Liège, 4000 Li\`ege, Belgium \\
$^3$Department of Applied Physics, Eindhoven University of Technology, P.O. Box 513, 5600 MB Eindhoven, The Netherlands}

\author{}
\affiliation{}

\date{\today}

\begin{abstract}

Understanding, manipulating, and using magnons – the quanta of spin waves – for energy-efficient applications is one of the primary goals of magnonics.
In this paper, we consider a synthetic antiferromagnet in which one of the ferromagnetic layers is driven by spin-orbit torque.
We find that under specific conditions for the magnitude of the spin-orbit torque and field, magnon pairs are spontaneously produced by quantum fluctuations in a way that is similar to Hawking pair production near black-hole horizons. 
One of the magnons is generated near the interface with the spacer layer in one of the magnetic layers of the synthetic antiferromagnet, while the other magnon is produced in the other magnetic layer. 
We compute the magnon current due to these spontaneously generated magnon pairs and estimate the temperature below which they should become observable. 
Additionally, we find that the magnons are entangled, which makes them interesting for future applications in quantum magnonics.

\end{abstract}

\maketitle

\section{Introduction}

Amidst the ongoing climate crisis and escalating energy demand, the need has been arising for future-proof and energy-efficient technologies. In the field of magnonics, the aim is to dramatically reduce energy loss in information technology by using spin waves \cite{EnergyEff:Chumak2015,EnergyEff:CSABA20171471,EnergyEff:Puebla2020}.

Spin waves are collective excitations in ordered magnetic systems.
Unlike electrons, spin waves do not exhibit Joule heating.
This makes them energy-efficient candidates to replace the electron in information technology.
Many promising advances have been made in magnonics and new challenges have been laid out by Ref. \cite{magnonicsroadmap}.
However, there is one persistent challenge: spin waves experience relatively large damping.
This is because, contrary to electrons, spin waves are not conserved particles, which limits their efficiency. 
Therefore, an amplifier is interesting, as it counterbalances the dissipative nature of spin waves.
One method of amplification has been explored in Refs. \cite{PhysRevApplied.18.064026,wang2023spin}. 
Here, Harms \textit{et al}. proposed a synthetic antiferromagnet with spin-orbit torque (SOT) as shown in Fig. \ref{fig:setup}.
A synthetic antiferromagnet is a magnetic multilayer in which two ferromagnets are coupled antiferromagnetically \cite{Duine2018}.
The set-up also contains an inhomogeneous magnetic field, and the SOT keeps one layer of the synthetic antiferromagnet stabilized against the direction of the magnetic field.
This configuration leads to enhanced reflection of incoming spin waves \cite{PhysRevApplied.18.064026,wang2023spin}.

For applications of magnonics, a classical description of the spin waves typically suffices. 
Recently, however, the nascent field of quantum magnonics has explored applications of magnons in quantum information science \cite{YUAN20221}.
Examples include the coupling between a magnon and a superconducting qubit \cite{QM:TABUCHI2016729,QM:lachance-quirion_entanglement-based_2020}, spin-qubit entanglement through magnon excitation \cite{QM:fukami_opportunities_2021}, magnon Bose-Einstein condensation \cite{QMBE:chumak_bose-einstein_2009,QMBE:demidov_magnon_2008,QMBE:demidov_observation_2008,QMBE:demidov_thermalization_2007,QMBE:demokritov_boseeinstein_2006,QMBE:kuroe_magnetic_2008}, and spin superfluidity \cite{QM:Bunkov2020,QMSF:bozhko_bogoliubov_2019,QMSF:bozhko_supercurrent_2016,QMSF:qaiumzadeh_spin_2017,QMSF:wimmer_spin_2019,QMSF:yuan_experimental_2018}, as well as magnon squeezing \cite{QM:Kamra,QM:wuhrer_theory_2022,QM:wuhrer_magnon_2023}.
In this article, we show that in the set-up of Fig. \ref{fig:setup}, magnon pairs may be spontaneously produced by quantum fluctuations. 
This pair production is akin to Hawking pair production near a black-hole horizon, which stems from particle-antiparticle coupling near the horizon. 
In our case, the antiparticle corresponds to a negative energy magnon.
We investigate the consequences of spontaneous pair production for the magnon current and the current-current correlations.
We find from the latter that the magnon pairs are entangled.
The interest in magnon-magnon entanglement has grown because of its application in quantum communication and its fundamental role in the transition from classical to quantum magnetism.
Several techniques have already been proposed to entangle magnons, such as using a microwave cavity \cite{MMECavity:ren_long-range_2022,MMECavity:golkar_magnon-magnon_2023,MMECavity:wu_remote_2021,MMECavity:zhang_quantum_2019} or through the squeezing properties of an antiferromagnetic lattice \cite{MMEaFM:yuan_enhancement_2020,MMEaFM:azimi_mousolou_magnon-magnon_2021,MMEaFM:kamra_antiferromagnetic_2019,MMEaFM:liu_tunable_2023}.
The proposal we put forward here has the benefits that it requires driving by direct, rather than alternating current, and in the transient regime requires no driving at all. 
Moreover, the entangled magnons are spatially separated, which facilitates their detection.

The remainder of this paper is structured as follows.
First, we introduce the model in Sec. \ref{sec:setup and Model}.
In Sec. \ref{sec:General Current expression}, we derive the current for a general scattering matrix that mixes creation and annihilation operators, which is the relevant situation for the set-up under consideration and gives rise to the production of pairs. 
Sec. \ref{sec:Scattering theory} contains the scattering calculation that leads to the reflection and transmission coefficients and, ultimately, the scattering matrix.
In Sec. \ref{sec:Spontaneous emission}, we compute various observables related to pair production, such as magnon current and current-current correlations.
We show that there is entanglement between magnons and a crossover temperature below which quantum effects become dominant.
We discuss the results and conclusion in Sec. \ref{sec:Discussion and Conclusion}.

\section{Set-up and Model}\label{sec:setup and Model}

\begin{figure}
    \centering

\tikzset{every picture/.style={line width=0.75pt}} 

\begin{tikzpicture}[x=0.45pt,y=0.45pt,yscale=-1,xscale=1]

\draw  [draw opacity=0][fill={rgb, 255:red, 245; green, 166; blue, 35 }  ,fill opacity=0.5 ] (513.97,185.94) -- (493.89,159.99) -- (524.22,106.68) -- (564.51,97.09) -- cycle ;
\draw  [draw opacity=0][fill={rgb, 255:red, 184; green, 233; blue, 134 }  ,fill opacity=1 ] (332.25,106.75) -- (359.13,106.75) -- (328.68,160.41) -- (301.8,160.41) -- cycle ;
\draw  [draw opacity=0][fill={rgb, 255:red, 80; green, 227; blue, 194 }  ,fill opacity=0.5 ] (136.52,160.41) -- (302.06,160.41) -- (302.06,195.69) -- (136.52,195.69) -- cycle ;
\draw  [draw opacity=0][fill={rgb, 255:red, 245; green, 166; blue, 35 }  ,fill opacity=0.5 ] (328.68,160.41) -- (494.23,160.41) -- (494.23,195.65) -- (328.68,195.65) -- cycle ;
\draw  [draw opacity=0][fill={rgb, 255:red, 245; green, 166; blue, 35 }  ,fill opacity=0.5 ] (358.76,107.08) -- (524.07,107.08) -- (493.99,160.41) -- (328.68,160.41) -- cycle ;
\draw  [draw opacity=0][fill={rgb, 255:red, 80; green, 227; blue, 194 }  ,fill opacity=0.5 ] (166.35,106.75) -- (332.25,106.75) -- (302.06,160.41) -- (136.16,160.41) -- cycle ;
\draw    (483.5,153.03) -- (502.97,115.29) ;
\draw [shift={(503.89,113.51)}, rotate = 117.29] [color={rgb, 255:red, 0; green, 0; blue, 0 }  ][line width=0.75]    (10.93,-3.29) .. controls (6.95,-1.4) and (3.31,-0.3) .. (0,0) .. controls (3.31,0.3) and (6.95,1.4) .. (10.93,3.29)   ;
\draw    (155.43,150.71) -- (174.48,116.76) ;
\draw [shift={(175.46,115.01)}, rotate = 119.3] [color={rgb, 255:red, 0; green, 0; blue, 0 }  ][line width=0.75]    (10.93,-3.29) .. controls (6.95,-1.4) and (3.31,-0.3) .. (0,0) .. controls (3.31,0.3) and (6.95,1.4) .. (10.93,3.29)   ;
\draw [color={rgb, 255:red, 74; green, 144; blue, 226 }  ,draw opacity=1 ]   (198.31,152.22) -- (217.36,118.27) ;
\draw [shift={(218.34,116.53)}, rotate = 119.3] [color={rgb, 255:red, 74; green, 144; blue, 226 }  ,draw opacity=1 ][line width=0.75]    (10.93,-3.29) .. controls (6.95,-1.4) and (3.31,-0.3) .. (0,0) .. controls (3.31,0.3) and (6.95,1.4) .. (10.93,3.29)   ;
\draw [color={rgb, 255:red, 208; green, 2; blue, 27 }  ,draw opacity=1 ]   (461.12,111.88) -- (441.64,149.62) ;
\draw [shift={(440.73,151.4)}, rotate = 297.29] [color={rgb, 255:red, 208; green, 2; blue, 27 }  ,draw opacity=1 ][line width=0.75]    (10.93,-3.29) .. controls (6.95,-1.4) and (3.31,-0.3) .. (0,0) .. controls (3.31,0.3) and (6.95,1.4) .. (10.93,3.29)   ;
\draw [color={rgb, 255:red, 74; green, 144; blue, 226 }  ,draw opacity=1 ]   (309.48,134.48) .. controls (303.19,145.31) and (300.06,123.92) .. (293.76,134.76) .. controls (287.46,145.6) and (284.33,124.2) .. (278.04,135.04) .. controls (274.74,138.43) and (271.63,135.36) .. (264.21,135.28) ;
\draw [shift={(262.31,135.32)}, rotate = 0.62] [color={rgb, 255:red, 74; green, 144; blue, 226 }  ,draw opacity=1 ][line width=0.75]    (9.84,-2.96) .. controls (6.25,-1.25) and (2.97,-0.27) .. (0,0) .. controls (2.97,0.27) and (6.25,1.26) .. (9.84,2.96)   ;
\draw [color={rgb, 255:red, 208; green, 2; blue, 27 }  ,draw opacity=1 ]   (346.18,135.9) .. controls (352.62,125.22) and (355.45,146.68) .. (361.9,135.99) .. controls (368.35,125.31) and (371.18,146.77) .. (377.63,136.09) .. controls (380.97,132.79) and (384.03,135.93) .. (391.46,136.19) ;
\draw [shift={(393.35,136.19)}, rotate = 182] [color={rgb, 255:red, 208; green, 2; blue, 27 }  ,draw opacity=1 ][line width=0.75]    (9.84,-2.96) .. controls (6.25,-1.25) and (2.97,-0.27) .. (0,0) .. controls (2.97,0.27) and (6.25,1.26) .. (9.84,2.96)   ;
\draw    (326.41,156.72) -- (309.8,156.72) ;
\draw [shift={(307.8,156.72)}, rotate = 360] [fill={rgb, 255:red, 0; green, 0; blue, 0 }  ][line width=0.08]  [draw opacity=0] (8.4,-2.1) -- (0,0) -- (8.4,2.1) -- cycle    ;
\draw    (326.41,156.72) -- (335.64,140.73) ;
\draw [shift={(336.63,139)}, rotate = 119.98] [fill={rgb, 255:red, 0; green, 0; blue, 0 }  ][line width=0.08]  [draw opacity=0] (7.2,-1.8) -- (0,0) -- (7.2,1.8) -- cycle    ;
\draw  [draw opacity=0][fill={rgb, 255:red, 184; green, 233; blue, 134 }  ,fill opacity=1 ] (302.02,160.41) -- (329.25,160.41) -- (329.25,195.79) -- (302.02,195.79) -- cycle ;
\draw  [draw opacity=0][fill={rgb, 255:red, 80; green, 227; blue, 194 }  ,fill opacity=0.4 ] (83.67,169.78) -- (88.94,169.78) -- (88.94,205.16) -- (83.67,205.16) -- cycle ;
\draw  [draw opacity=0][fill={rgb, 255:red, 80; green, 227; blue, 194 }  ,fill opacity=0.3 ] (71.39,170.47) -- (76.66,170.47) -- (76.66,205.37) -- (71.39,205.37) -- cycle ;
\draw  [draw opacity=0][fill={rgb, 255:red, 80; green, 227; blue, 194 }  ,fill opacity=0.2 ] (59.11,169.43) -- (64.38,169.43) -- (64.38,205.37) -- (59.11,205.37) -- cycle ;
\draw  [draw opacity=0][fill={rgb, 255:red, 80; green, 227; blue, 194 }  ,fill opacity=0.1 ] (47.46,169.71) -- (52.72,169.71) -- (52.72,205.72) -- (47.46,205.72) -- cycle ;
\draw  [draw opacity=0][fill={rgb, 255:red, 80; green, 227; blue, 194 }  ,fill opacity=0.4 ] (133.86,79.98) -- (139.1,79.98) -- (88.93,169.78) -- (83.69,169.78) -- cycle ;
\draw  [draw opacity=0][fill={rgb, 255:red, 80; green, 227; blue, 194 }  ,fill opacity=0.3 ] (121.96,79.98) -- (127.18,79.98) -- (76.63,170.47) -- (71.41,170.47) -- cycle ;
\draw  [draw opacity=0][fill={rgb, 255:red, 80; green, 227; blue, 194 }  ,fill opacity=0.2 ] (109.1,79.98) -- (114.35,79.98) -- (64.37,169.43) -- (59.13,169.43) -- cycle ;
\draw  [draw opacity=0][fill={rgb, 255:red, 80; green, 227; blue, 194 }  ,fill opacity=0.1 ] (97.41,80.32) -- (102.07,80.32) -- (52.13,169.71) -- (47.47,169.71) -- cycle ;
\draw  [draw opacity=0][fill={rgb, 255:red, 80; green, 227; blue, 194 }  ,fill opacity=0.5 ] (146.26,80.68) -- (166.39,106.67) -- (136.27,160.83) -- (95.84,171.35) -- cycle ;
\draw  [draw opacity=0][fill={rgb, 255:red, 80; green, 227; blue, 194 }  ,fill opacity=0.5 ] (136.67,160.78) -- (136.51,195.31) -- (95.67,205.88) -- (95.84,171.35) -- cycle ;
\draw    (136.16,160.41) -- (136.51,195.31) ;
\draw    (166.35,106.75) -- (136.16,160.41) ;
\draw  [draw opacity=0][fill={rgb, 255:red, 245; green, 166; blue, 35 }  ,fill opacity=0.4 ] (527.17,220.17) -- (521.9,220.17) -- (521.9,186.21) -- (527.17,186.21) -- cycle ;
\draw  [draw opacity=0][fill={rgb, 255:red, 245; green, 166; blue, 35 }  ,fill opacity=0.3 ] (539.45,220.29) -- (534.19,220.29) -- (534.19,186) -- (539.45,186) -- cycle ;
\draw  [draw opacity=0][fill={rgb, 255:red, 245; green, 166; blue, 35 }  ,fill opacity=0.2 ] (551.73,220.5) -- (546.47,220.5) -- (546.47,186) -- (551.73,186) -- cycle ;
\draw  [draw opacity=0][fill={rgb, 255:red, 245; green, 166; blue, 35 }  ,fill opacity=0.1 ] (563.38,220.23) -- (558.12,220.23) -- (558.12,185.67) -- (563.38,185.67) -- cycle ;
\draw  [draw opacity=0][fill={rgb, 255:red, 245; green, 166; blue, 35 }  ,fill opacity=0.4 ] (527.16,186.21) -- (521.92,186.21) -- (572.09,96.4) -- (577.33,96.4) -- cycle ;
\draw  [draw opacity=0][fill={rgb, 255:red, 245; green, 166; blue, 35 }  ,fill opacity=0.3 ] (539.42,186) -- (534.2,186) -- (584.76,95.5) -- (589.98,95.5) -- cycle ;
\draw  [draw opacity=0][fill={rgb, 255:red, 245; green, 166; blue, 35 }  ,fill opacity=0.2 ] (551.72,186) -- (546.48,186) -- (596.46,96.54) -- (601.7,96.54) -- cycle ;
\draw  [draw opacity=0][fill={rgb, 255:red, 245; green, 166; blue, 35 }  ,fill opacity=0.1 ] (563.6,185.67) -- (558.13,185.67) -- (608.03,96.36) -- (613.49,96.36) -- cycle ;
\draw  [draw opacity=0][fill={rgb, 255:red, 245; green, 166; blue, 35 }  ,fill opacity=0.5 ] (494.23,195.65) -- (494.39,160.62) -- (513.97,185.94) -- (513.8,220.97) -- cycle ;
\draw    (493.97,160.08) -- (494.23,195.65) ;
\draw    (524.12,107.06) -- (493.94,160.72) ;

\draw (139.95,171.24) node [anchor=north west][inner sep=0.75pt]  [font=\small] [align=left] {FM};
\draw (458.28,126.45) node [anchor=north west][inner sep=0.75pt]    {$h_{R}$};
\draw (168.75,126.45) node [anchor=north west][inner sep=0.75pt]    {$h_{L}$};
\draw (450.28,171.24) node [anchor=north west][inner sep=0.75pt]  [font=\small] [align=left] {FM};
\draw (331.77,123.84) node [anchor=north west][inner sep=0.75pt]  [font=\footnotesize]  {$z$};
\draw (313.82,139.64) node [anchor=north west][inner sep=0.75pt]  [font=\footnotesize]  {$x$};
\draw (256.77,169.26) node [anchor=north west][inner sep=0.75pt]   [align=left] {Left};
\draw (333.69,169.26) node [anchor=north west][inner sep=0.75pt]   [align=left] {Right};
\draw (400.58,124.08) node [anchor=north west][inner sep=0.75pt]  [color={rgb, 255:red, 208; green, 2; blue, 27 }  ,opacity=1 ]  {$\vec{m}_{R}$};
\draw (215.55,124.08) node [anchor=north west][inner sep=0.75pt]  [color={rgb, 255:red, 74; green, 144; blue, 226 }  ,opacity=1 ]  {$\vec{m}_{L}$};
\draw (158.02,70.85) node [anchor=north west][inner sep=0.75pt]    {$\mu _{L} ,T_{L} \ $};
\draw (520.84,70.83) node [anchor=north west][inner sep=0.75pt]    {$\mu _{R} ,T_{R} \ $};

\end{tikzpicture}

    \caption{Schematic of the synthetic antiferromagnet. The left and right ferromagnets are colored blue and red, respectively. The ferromagnets are coupled to a heavy metal layer that induces spin-orbit torque. This effect is modeled through the leads at fixed temperatures $T_{L/R}$ and spin accumulation $\mu_{L/R}$. The left ferromagnet magnetization $\vec m_L$ points in the direction of the left magnetic field $h_L \hat z$. The right ferromagnet has its magnetization $\vec m_R$ opposite to the right magnetic field $h_R \hat z$. The alignment opposite to the magnetic field is achieved by injecting angular momentum through spin-orbit torque. The nonmagnetic layer in green couples the ferromagnets antiferromagnetically. The left magnons couple to the negative energy magnons on the right, which ultimately leads to the creation of pairs of entangled magnons, as indicated by the wiggly lines.}
    \label{fig:setup}
\end{figure}

In this section, we discuss the set-up and model.
We consider a synthetic antiferromagnet as is shown in Fig. \ref{fig:setup}. 
It consists of two ferromagnets (FMs) coupled antiferromagnetically, and where a metallic reservoir separately contacts each ferromagnet.
Antiferromagnetic coupling is achieved by placing a nonmagnetic layer between the FMs.
The interlayer exchange coupling is a Ruderman-Kittel-Kasuya-Yosida (RKKY) coupling that spatially oscillates \cite{RKKY1,RKKY2}.
As a result, the thickness of the nonmagnetic layer is chosen such that the two FMs, which we refer to as left and right FMs, are coupled antiferromagnetically.
This configuration is then referred to as a synthetic antiferromagnet \cite{Duine2018}.
The left and right FM have a magnetization $\vec m_{L/R}$ and the left FM magnetization aligns with the magnetic field $h_L \vec z$.
The right FM magnetization aligns opposite to the magnetic field $h_R \vec z$.
FM spins opposite to the magnetic field are energetically unstable; thus, the spins will relax to align with the magnetic field if there is no driving and if the antiferromagnetic coupling is weak.
To prevent this relaxation, we introduce a spin current into the ferromagnetic layer, injecting angular momentum to counteract the loss caused by damping.
In practice, this driving is achieved through a heavy metal layer that induces a spin-orbit torque (SOT): an electrical current in the heavy metal that is tangential to the interface between the heavy metal and the FM leads, via the spin Hall effect, to a spin current through the interface. 
This ensures that the right FM is dynamically stabilized \cite{SOTManchon2015, SOTMiron2011, SOTPhysRevLett.109.096602, SOTRevModPhys.91.035004}. 
The spin Hall effect gives rise to a nonzero spin accumulation in the heavy metal, which as a result acts as a spin reservoir for the FMs \cite{ChemicalPot}. This spin accumulation is denoted by $\mu_L$ and $\mu_R$ for the left and right reservoir, respectively. 
For the spins in the right FM to be oppositely aligned with the magnetic field it is required that $\mu_R<-h_R$ \cite{antimagnonics}.
In Fig. 1, we have fixed the temperatures $T_L$ and $T_R$. 
The left and right FMs are modeled by the Hamiltonians

\begin{align}
    \hat H_L=&-J\sum_{\langle i,j\rangle}\vec S^L_{i}\cdot \vec S^L_{j}- h_L\sum_i  \hat S^L_{z,i}, \\
    \hat H_R=&-J\sum_{\langle i,j\rangle}\vec S^R_i\cdot \vec S^R_{j}- h_R\sum_i  \hat S^R_{z,i}.
\end{align}

\noindent Here $J$ is the exchange coupling within each FM that we assume, for simplicity, to be equal for both FMs.
For simplicity, we neglect the contribution of anisotropies.
However, these could be reinstated straightforwardly.
To simplify the computations, we take the system to be one-dimensional. 
Generalization to higher dimensions is straightforward as well. 
First, we perform the Holstein-Primakoff transformations with $\vec z$ as the quantization axis

\begin{align*}
    \hat S^{L}_{+,i} &= \hbar\sqrt{2S-\psid_{Li} \hat \psi_L}\hat \psi_{Li}, \\
    \hat S^{L}_{-,i} &= \hbar\psid_L\sqrt{2S-\psid_{Li} \hat \psi_{Li}}, \\
    \hat S^{L}_{z,i}&=\hbar (S-\hat n_{Li}).
\end{align*}

\noindent Here, $S$ is the total spin.
The bosonic annihilation (creation) operators $\hat \psi^{(\dagger)}_{Li}$ annihilate (create) a quanta of spin wave, magnon, on site $i$ in the left FM.
The number operator is defined as $\hat n_{Li}=\psid_{Li} \hat \psi_{Li}$.
Inserting the Holstein-Primakoff transformation and expanding up to quadratic order yields the Hamiltonian for the left FM

\begin{align}
    \hat H_L&=-\frac{\hbar JS}{2}\sum_{i\in U_L}\sum_{j=\pm a} \{\psid_{L i}\hat \psi_{L i+j}+\hat \psi_{L i}\psid_{L i+j}\} \nonumber \\
    &+ \hbar (2JS+h_L) \sum_{i\in U_L} \hat n_{L i}+E_0^L.
\end{align}

\noindent Here, $U_{L}$ are the lattice positions on the left with lattice spacing $a$ and $E_0^L$ is the ground energy of the left FM.
For the right FM, we stabilize the spins opposite to the magnetic field with SOT.
Therefore, we need a Holstein-Primakoff transformation that uses the $-\vec z$ axis

\begin{align*}
    \hat S^{R}_{+,i} &= \hbar\psid_{Ri}\sqrt{2S-\psid_{Ri} \hat \psi_{Ri}}, \\
    \hat S^{R}_{-,i} &= \hbar\sqrt{2S-\psid_{Ri} \hat \psi_{Ri}}\hat \psi_{Ri},\\
    \hat S^{R}_{z,i} &=\hbar (\hat n_{Ri}-S) .
\end{align*}

\noindent As before, $S$ is the total spin that we consider to be equal for both layers.
The bosonic annihilation (creation) operators $\hat \psi^{(\dagger)}_{Ri}$ annihilate (create) a quanta of a spin wave, magnon, on site $i$ in the right FM.
The magnon number operator is defined as $\hat n_{Ri}=\psid_{Ri} \hat\psi_{Ri}$.
We find that

\begin{align}
    \hat H_R&=-\frac{\hbar JS}{2}\sum_{i\in U_R}\sum_{j=\pm a} \{\psid_{R i}\hat \psi_{R i+j}+\hat \psi_{R i}\psid_{R i+j}\} \nonumber\\
    &+\hbar (2JS-h_R) \sum_{i\in U_R} \hat n_{R i}+E_0^R.
\end{align}

\noindent Here, $U_{R}$ are the lattice positions in the right FM with lattice spacing $a$, and $E_0^R$ is the energy of the right FM given that its spins are opposite to the magnetic field.
For both the left and right FM, we take the continuum limit, which results in the Hamiltonian

\begin{align}\label{eq:nonH}
    \hat H_\nu=&\frac{\hbar }{2}\int dx \hspace{3px}\psid_\nu(x,t)(-JSa^2\partial_x^2+\sigma h_\nu)\hat \psi_\nu(x,t)  +h.c.
\end{align}

\noindent here, $\nu\in\{L,R\}$ and the integral is over the left/right volume.
The sign in front of $h_\nu$ is given by $\sigma$.
Here $\sigma=+1$ for the left magnet and $\sigma=-1$ for the right magnet.
The operator $\psi^{(\dagger)}_\nu(x,t)$ annihilates (creates) an excitation of the spin in the left or right FM for $\nu=L$ or $\nu=R$, respectively.
We work in natural units for the remainder of this paper, i.e. $\hbar=1$.

\subsection{Boundary conditions}

The FMs are coupled antiferromagnetically through the RKKY coupling.
This interaction is implemented through the additional term

\begin{equation} \label{eq:boundary}
    \hat H_{\text{RKKY}}=J'\Vec{S}_L(0)\cdot\Vec{S}_{R}(0).
\end{equation}

\noindent Recall that $x=0$ is the boundary where the left and right FMs meet.
We have that $J'>0$.
Incorporating this term allows us to derive the constraints of the magnon operators at $x=0$.

First, we substitute the Holstein-Primakoff transformation for $\Vec{S}_L(0)$ and $\Vec{S}_{R}(0)$.
After linearization, we find from Eq. (\ref{eq:boundary}) that

\begin{align}
    \hat H_{\text{RKKY}}=\frac{J'S}{2}\big[&\psid_L(0)\hat \psi_L(0)+\psid_R(0)\hat\psi_R(0) \nonumber\\
    &+\psid_L(0) \psid_R(0)+\hat \psi_L(0)\hat\psi_R(0)\big].
\end{align}

\noindent This additional term imposes the following conditions on the operators when computing their Heisenberg equations of motion

\begin{align} 
\partial_{x}\hat \psi_{L}(0,t)-\gamma(\hat \psi_L(0,t)+\psid_R(0,t))&=0, \label{eq:BEOM1} \\
\partial_{x}\hat \psi_R(0,t)+\gamma(\hat \psi_R(0,t)+\psid_L(0,t))&=0,\label{eq:BEOM2}
\end{align}

\noindent with $\gamma=-J'/2Ja$.
These boundary conditions are necessary for computing the scattering coefficients of the scattering states, as we will see in Sec. \ref{sec:Scattering theory}.

\section{General Current expression}\label{sec:General Current expression}

The aim of the sections hereafter is to compute the scattering matrix for the magnons and the observables that can be computed from it.
One of those observables is the expectation value of the current.
Therefore, we derive in this section a general expression for the current.
Conventionally, in set-ups involving reservoirs such as our set-up in Fig. 1, and for sufficiently small systems such that the transport is phase-coherent, the current is computed using the Landauer-B\"uttiker formalism \cite{BLANTER20001}.
Unfortunately, this will not suffice for our purposes, since this formalism does not mix creation and annihilation operators.
The driven synthetic antiferromagnet of Fig.\ref{fig:setup} allows for positive and negative energy magnons.
For sufficiently large wavelengths, a magnon in the right FM lowers the total energy of the system.
Thus, such a magnon carries negative energy.
This reflects the tendency of the spins in the right FM to align with the magnetic field.
The positive-energy magnons in the left FM and negative-energy magnons in the right FM may couple since a positive-energy magnon on the left precesses in the same direction as a negative-energy magnon on the right.
Due to the conservation of energy, an excitation of a positive energy magnon has to be paired with an excitation of a negative energy magnon.
To account for this pair-production process, the creation and annihilation operators of the positive- and negative-energy magnons have to mix in the scattering matrix.
Therefore we need to revisit the Landauer-B\"uttiker formalism to take this into account.

Consider the setup as in Fig. \ref{fig:setup}.
We describe the system using a basis of bosonic operators that annihilate (create) ingoing magnons $\{\hat a^{\text{in}}_i(\omega)\}$ and bosonic operators that annihilate (create) outgoing magnons $\{\hat a^{\text{out}}_i(\omega)\}$.
The index $i$ is $L$ for the positive energy magnon on the left and $R$ for the negative energy magnon on the right.
Then the scattering matrix relates the bases as

\begin{equation}\label{eq:GenSmatrix}
    \begin{pmatrix}
        \hat a^{\text{out}}_{L}(\omega) \\
        (\hat a^{\text{out}}_{R}(-\omega))^\dagger 
    \end{pmatrix}=S
    \begin{pmatrix}
        \hat a^{\text{in}}_{L}(\omega) \\
        (\hat a^{\text{in}}_{R}(-\omega))^\dagger 
    \end{pmatrix},
\end{equation}

\noindent where $S$ is given by

\begin{equation}
    S=
    \begin{pmatrix}
        R_{L}(\omega) & T_L(\omega) \\
        T_R(\omega) & R_{R}(\omega)
    \end{pmatrix}.
\end{equation}

\noindent $R_{L/R}(\omega)$ are the reflection coefficients on either side of the interface and $T_{L/R}(\omega)$ is the transmission coefficient which is non-zero for frequencies where the positive and negative energy magnons overlap. 
The $S$-matrix satisfies the following condition, $S\eta S^{\dagger}=\eta$ with $\eta=\text{diag}(1,-1)$.
This condition is due to the conservation of the Bogoliubov norm of the magnons, as is shown in the next section and in Ref. \cite{AG:coutant_black_2012,AG:larre_quantum_2012}.
The current has the form \cite{BLANTER20001}

\begin{align}\label{eq:currentE}
    \hat J_{i}(t)=&\int d\omega d\omega'((\hat a^{\text{in}}_{i} (\omega))^\dagger\hat a^{\text{in}}_{i}(\omega')-(\hat a^{\text{out}}_{i}(\omega))^\dagger\hat a^{\text{out}}_{i}(\omega')) \nonumber\\
    &\times e^{i(\omega-\omega')t/\hbar}. 
\end{align}

\noindent This result together with the relation between in- and outgoing scattering operators in Eq. (\ref{eq:GenSmatrix}) gives the spin current

\begin{align}
    \langle\hat J_{i}\rangle=& -\int d\omega |T(\omega)|^2(1+f_{i}(\omega)+f_{j}(-\omega)), \label{LBform}
\end{align}

\noindent with $i\neq j$.
Here, we used that $\langle (\hat a^{\text{in}}_{i}(\omega) )^\dagger\hat a^{\text{in}}_{j}(\omega')\rangle=f_{i}(\omega)\delta_{ij}\delta(\omega-\omega')$, where $f_i(\omega)$ is the distribution of lead $i$.
The current above gives the flow of magnons and the sign is chosen such that a positive current implies magnons flowing away from the reservoir.
Both the left and right magnon currents carry a minus sign, which indicates that magnons are flowing from the interface into the reservoirs.
The flow of spin angular momentum is computed by incorporating the sign of the group velocity of the magnon.
This provides an extra minus sign for $\langle \hat J_R\rangle$, therefore, the flow of spin angular momentum is always from right to left.
This is due to the field on the right being larger. 
Eq. (\ref{LBform}) has the interesting property that even when the distribution functions vanish, there is a current that persists.
This is the spontaneous pair production of magnons.
In Appendix \ref{noSOT}, we contrast this result with a synthetic antiferromagnet in its ground state.
Then, we retrieve the conventional result known from Landauer-B\"uttiker formalism.

We note that the spontaneous pair production arises as long as there is a frequency window where the positive- and negative-energy magnons couple across the interface between the left and right FM. 
For an expression of the current that is valid for all frequencies, we refer to Eq. (\ref{LBformAp}).
In the next section, we explicitly compute the scattering matrix and discuss the conditions where such a frequency window exists.

\section{$S$-matrix}\label{sec:Scattering theory}

The scattering matrix allows us to compute the number of outgoing magnons and, more specifically, the transmission coefficient associated with spontaneous pair production in Eq. (\ref{LBform}).
We now compute the scattering matrix explicitly, and show that the intermediate result is equivalent to the enhanced reflection derived in Ref. \cite{PhysRevApplied.18.064026}.
The computational steps we use are found in several articles on the flow of Bose-Einstein condensates with discontinuities \cite{BECPhysRevA.80.043603, BECPhysRevD.83.124047, BECPhysRevD.83.124016}, albeit that our Hamiltonian is different. 

\begin{figure}
    \centering

\tikzset{every picture/.style={line width=0.75pt}} 

\begin{tikzpicture}[x=0.55pt,y=0.6pt,yscale=-1,xscale=1]

\draw  (64,148.6) -- (248.51,148.6)(156.09,78.78) -- (156.09,217.42) (241.51,143.6) -- (248.51,148.6) -- (241.51,153.6) (151.09,85.78) -- (156.09,78.78) -- (161.09,85.78)  ;
\draw  [color={rgb, 255:red, 74; green, 144; blue, 226 }  ,draw opacity=1 ] (94.34,79.42) .. controls (136.17,158.13) and (178,158.13) .. (219.84,79.42) ;
\draw  [color={rgb, 255:red, 208; green, 2; blue, 27 }  ,draw opacity=1 ] (218.94,214.42) .. controls (177.1,149.11) and (135.27,149.11) .. (93.44,214.42) ;
\draw  [color={rgb, 255:red, 74; green, 144; blue, 226 }  ,draw opacity=1 ] (305.75,79.42) .. controls (360.88,223.59) and (416.01,223.59) .. (471.14,79.42) ;
\draw  [dash pattern={on 4.5pt off 4.5pt}]  (131.79,128.91) -- (453.33,129.2) ;
\draw [color={rgb, 255:red, 74; green, 144; blue, 226 }  ,draw opacity=1 ] [dash pattern={on 4.5pt off 4.5pt}]  (181.98,128.91) -- (182.33,149.86) ;
\draw [color={rgb, 255:red, 74; green, 144; blue, 226 }  ,draw opacity=1 ] [dash pattern={on 4.5pt off 4.5pt}]  (131.79,128.91) -- (132.03,151.43) ;
\draw [color={rgb, 255:red, 74; green, 144; blue, 226 }  ,draw opacity=1 ] [dash pattern={on 4.5pt off 4.5pt}]  (328.72,130.28) -- (328.95,152.8) ;
\draw [color={rgb, 255:red, 74; green, 144; blue, 226 }  ,draw opacity=1 ] [dash pattern={on 4.5pt off 4.5pt}]  (448.83,130.22) -- (449.06,152.75) ;
\draw [color={rgb, 255:red, 208; green, 2; blue, 27 }  ,draw opacity=1 ] [dash pattern={on 4.5pt off 4.5pt}]  (352.76,130.22) -- (352.99,152.75) ;
\draw [color={rgb, 255:red, 208; green, 2; blue, 27 }  ,draw opacity=1 ] [dash pattern={on 4.5pt off 4.5pt}]  (425.19,129.91) -- (425.43,152.43) ;
\draw  [color={rgb, 255:red, 208; green, 2; blue, 27 }  ,draw opacity=1 ] (471.14,217.55) .. controls (416.01,73.39) and (360.88,73.39) .. (305.75,217.55) ;
\draw  [dash pattern={on 4.5pt off 4.5pt}]  (102,138.2) -- (157.09,138.45) ;
\draw  [dash pattern={on 4.5pt off 4.5pt}]  (388.45,109.42) -- (476,108.86) ;
\draw  (296.27,148.76) -- (480.79,148.76)(388.36,78.94) -- (388.36,217.58) (473.79,143.76) -- (480.79,148.76) -- (473.79,153.76) (383.36,85.94) -- (388.36,78.94) -- (393.36,85.94)  ;

\draw (168.77,69.04) node [anchor=north west][inner sep=0.75pt]    {$\omega $};
\draw (397.96,69.59) node [anchor=north west][inner sep=0.75pt]    {$\omega $};
\draw (251.58,139.74) node [anchor=north west][inner sep=0.75pt]    {$k$};
\draw (486.8,139.67) node [anchor=north west][inner sep=0.75pt]    {$k$};
\draw (184.05,153.67) node [anchor=north west][inner sep=0.75pt]    {$k_{r}^{L}$};
\draw (113.6,152.87) node [anchor=north west][inner sep=0.75pt]    {$k_{l}^{L}$};
\draw (452.27,151.03) node [anchor=north west][inner sep=0.75pt]    {$k_{r}^{R}$};
\draw (306.27,152.37) node [anchor=north west][inner sep=0.75pt]    {$k_{l}^{R}$};
\draw (400.96,150.79) node [anchor=north west][inner sep=0.75pt]    {$k_{+}^{R}$};
\draw (355.33,151.15) node [anchor=north west][inner sep=0.75pt]    {$k_{-}^{R}$};
\draw (79.33,127.4) node [anchor=north west][inner sep=0.75pt]    {$h_{L}$};
\draw (479.67,98.73) node [anchor=north west][inner sep=0.75pt]    {$h_{R}$};

\end{tikzpicture}

    \caption{In the left plot, we see the dispersion relation in the left FM. Blue are positive frequency excitations, and red are negative frequency excitations. The right plot represents the right FM. Driving and the inhomogeneous field causes dispersions to overlap, resulting in multiple modes that couple due to scattering processes.}
    \label{fig:dispersions}
\end{figure}
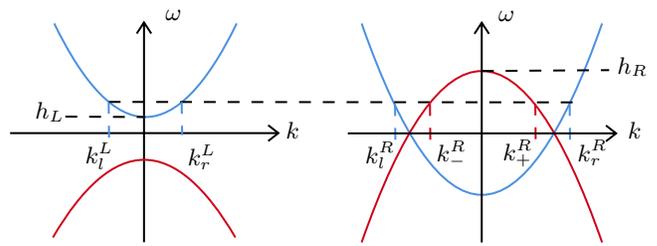

The Fourier transforms of the operators are as follows

\begin{equation}
    \hat \psi_\nu(x,t)=\int \frac{dk}{2\pi} \hat \psi_\nu(k,t)e^{ikx}. 
\end{equation}

\noindent This is substituted into the non-interacting Hamiltonian Eq. (\ref{eq:nonH})

\begin{align}
    \hat H=&\int \frac{dk}{2\pi}\omega_k^\nu\psid_{\nu}(k,t)\hat \psi_{\nu}(k,t).
\end{align}

\noindent The frequencies $\omega^L_{k}$ and $\omega^R_{k}$ are given by

    \begin{align}
    \omega^\nu_{k}&=Jsa^2k^2+ \sigma h_{\nu}.
\end{align}

\noindent The dispersion relation allows us to match plane wave solutions since only spin waves with equal frequency can couple.
First, we need to decompose the magnon operator to allow for negative frequencies, equivalent to counterclockwise-precessing spins.
We write

\small

\begin{align} \label{eq:decomposition}
    \hat \psi_{\nu}(x,t) &=\int_{-\infty}^\infty d\omega \hspace{2px} \hat a(\omega)\phi_{\nu}(\omega,x) e^{i\omega t}+ (\hat a(\omega))^\dagger (\varphi_{\nu}(\omega,x))^* e^{-i\omega t}.
\end{align}
\normalsize

\noindent The operators $\hat a(\omega)$ and $(\hat a(\omega))^\dagger$ are magnon annihilation and creation operators that satisfy the bosonic commutation rules $[\hat a(\omega),(\hat a(\omega))^\dagger]=1$.
The positive and negative frequency modes at $\omega$ are denoted by $\phi_\nu(\omega,x)$ and $\varphi_\nu(\omega,x)$, respectively. 
This decomposition allows for both clockwise precession and counterclockwise precession of spins.
Counter-clockwise precession can be neglected in most cases since it does not contribute in an energetically stable configuration.
However, in our setup, the counterclockwise precessing spin waves on spins oppositely aligned to the magnetic field contribute non-trivially to the dynamics.

We plot the dispersion relations associated with $\phi_\nu(\omega,x)$ (blue) and $\varphi_\nu(\omega,x)$ (red) in Fig. \ref{fig:dispersions}.
The plot on the left shows the dispersion in the left FM; only one of the dispersions is positive and will contribute to the scattering process.
In the right plot, we have the dispersion relation of the right FM.
Both dispersions enter the positive frequency region, implying that another set of spin wave solutions contributes to the scattering process.
The red plot represents the counter-precessing modes and if the dispersion enters the positive frequency region then they can be identified as negative energy excitations (with energy $-\hbar \omega$).
We consider now a scattering process with an incoming magnon in the left FM at frequency $\omega$, with $h_L < \omega < h_R$. 
The wavevectors as in Fig. \ref{fig:dispersions} are given by

\begin{subequations}
\begin{align}
    \Lambda k^L_{r}&=\sqrt{\omega -h_l}, & \Lambda k^L_{+}&= i\sqrt{\omega +h_l}, \label{mom1} \\ 
    \Lambda k^L_{l}&=-\sqrt{\omega -h_l}, & \Lambda k^L_{-}&=- i\sqrt{\omega +h_l}, \\ 
    \Lambda k^R_{r}&=\sqrt{\omega +h_R}, & \Lambda k^R_{+}&=\sqrt{h_R-\omega}, \\
    \Lambda k^R_{l}&=-\sqrt{\omega +h_R}, & \Lambda k^R_{-}&=-\sqrt{h_R-\omega}. \label{mom2}
\end{align}
\end{subequations}

\noindent Here, $\Lambda^2={JS}a^2$.
In the left ferromagnet, $k^L_{r/l}$ is the wave vectors of the magnon moving right and left, respectively.
Furthermore, $k^L_{+/-}$ describes the exponential damping and growing mode. 
In the right FM, $k^R_{r/l}$ correspond to the right and left-moving positive-frequency magnon modes.
Similarly, $k^R_{+/-}$ are the wave vectors of the negative frequency magnon modes moving to the right and left, respectively.

Recall that the Holstein-Primakoff operator is decomposed into positive and negative frequency precessing modes, Eq. (\ref{eq:decomposition}).
We express these modes as a sum of travelling spin waves.
These spin waves are matched at the boundary using the boundary conditions in Eq. (\ref{eq:BEOM1}-\ref{eq:BEOM2}), which allows us to compute reflection and transmission coefficients.
Thus $\phi_{\nu}(\omega,x)$ and $\varphi_{\nu}(\omega,x)$ are written as a linear combination of travelling waves:

\begin{subequations}
    
\begin{align} \label{aLi1}
    \phi_\nu(\omega,x)=&A_{r}^{\nu}u_{r}^{\nu}e^{ik^\nu_r(\omega)x}+A_{l}^{\nu}u_{l}^{\nu}e^{ik^\nu_l(\omega)x}  \nonumber\\ 
    +&A_{+}^{\nu}u_{+}^{\nu}e^{ik^\nu_+(\omega)x}+A_{-}^{\nu}u_{-}^{\nu}e^{ik^\nu_-(\omega)x}, \\\label{aLi2}
 \varphi_\nu(\omega,x)=&A_{r}^{\nu}v_{r}^{\nu}e^{ik^\nu_r(\omega)x} +A_{l}^{\nu}v_{l}^{\nu}e^{ik^\nu_l(\omega)x} \nonumber\\ 
    +&A_{+}^{\nu}v_{+}^{\nu}e^{ik^\nu_+(\omega)x} +A_{-}^{\nu}v_{-}^{\nu}e^{ik^\nu_-(\omega)x}. 
\end{align} 
\end{subequations}

\noindent Here, $A_{i}^{\nu}$ are the amplitudes of each magnon mode where $i$ sums over the four possible wave numbers for a given $\omega$, $i\in\{l,r,+,-\}$ as given in Eq. (\ref{mom1}-\ref{mom2}).
Furthermore, $u_{i}^{\nu}$ and $v_{i}^{\nu}$ are Bogoliubov coefficients that satisfy $|u_{i}^{\nu}|^2-|v_{i}^{\nu}|^2=1$.
The decomposition into travelling waves does not cover the uniform mode ($k=0$).
The uniform mode requires a separate analysis that includes non-linear effects, which is beyond the scope of this paper.
In the set-up, we only have circular spin waves, which results in Bogoliubov modes with values $u_{l/r}^{\nu}=1$, $v_{+/-}^{\nu}=1$, while the modes  $u_{+/-}^{\nu}$, $v_{l/r}^{\nu}$ remain zero.
For elliptical spin waves, the Bogoliubov coefficients are all non-zero. 
This complicates the calculation without yielding a qualitative difference.

We are now in the position to compute the scattering coefficients, but first, we need to define the scattering states.
This allows us to construct the scattering matrix.
The ingoing scattering state is composed of a magnon that travels towards the interface between the FMs from $\pm\infty$ and multiple magnons leaving from either side of the interface.
The outgoing scattering state is composed of a magnon that travels away from the interface towards $\pm\infty$ and multiple magnons propagating towards the interface in both FMs.
The in- and out-going scattering states are given by

\begin{subequations}
\begin{align}
    \phi^{\text{in}}_L(x)=& 
\begin{cases}
    e^{ik^L_rx}+Re^{ik^L_lx}, & \text{if } x<0\\
    Te^{ik^R_+x}+\Tilde T'e^{ik^R_rx},              & \text{otherwise,}
\end{cases} \\
    \phi^{\text{in}}_{Rp}(x)=& 
\begin{cases}
    \Tilde{T}e^{ik^L_lx}, & \text{if } x<0\\
    e^{ik^R_lx}+\Tilde R e^{ik^R_rx}, & \text{otherwise,}
\end{cases} \\
\phi^{\text{in}}_{Rn}(x)=& 
\begin{cases}
    T'e^{ik^L_lx}, & \text{if } x<0\\
    e^{ik^R_-x}+R'e^{ik^R_+x}, & \text{otherwise,}
\end{cases} 
\end{align}
\end{subequations}
\begin{subequations}
\begin{align}
\phi^{\text{out}}_L(x)= &
\begin{cases}
    e^{ik^L_lx}+R^*e^{ik^L_rx}, & \text{if } x<0\\
    T^*e^{ik^R_-x}+(\Tilde T')^*e^{ik^R_lx},              & \text{otherwise,}
\end{cases}\\
    \phi^{\text{out}}_{Rp}(x)=& 
\begin{cases}
    (\Tilde T)^*e^{ik^L_rx}, & \text{if } x<0\\
    e^{ik^R_rx}+ (\Tilde R)^*e^{ik^R_lx}, & \text{otherwise,}
\end{cases} \\
\phi^{\text{out}}_{Rn}(x)=& 
\begin{cases}
    (T')^*e^{ik^L_rx}, & \text{if } x<0\\
    e^{ik^R_+x}+(R')^*e^{ik^R_-x}, & \text{otherwise.}
\end{cases} 
\end{align}
\end{subequations}

\noindent The in/out index describes a scattering state with a single magnon going in/out from the left (L) or right (R) FM and the $p$ ($n$) index signals if the in or outgoing magnon has a positive (negative) frequency.
The $T$, $T'$, $\Tilde T$, and $\Tilde T'$ are transmission coefficients.
The $R$, $R'$, and $\Tilde R$ are reflection coefficients.
We have omitted their dependence of $\omega$ for brevity.
From now on we refer to in- and out-going scattering states as in- and out-states.
We expand our field operator into the in- and out-state bases

\begin{subequations}

\begin{align}
    \hat \psi(x,t) =\int_{h_L}^{h_R} d\omega \hspace{3px}        &\phi^{\text{in}}_L\hat a^{\text{in}}_L(\omega)e^{i\omega t} + \phi^{\text{in}}_{Rp}\hat a^{\text{in}}_{Rp}(\omega)e^{i\omega t} \nonumber\\  \label{eq:decomp1}
    &+ \phi^{\text{in}}_{Rn}(\hat a^{\text{in}}_{Rn}(-\omega))^\dagger e^{-i\omega t}, \\
    \hat \psi(x,t) =\int_{h_L}^{h_R} d\omega \hspace{3px}        &\phi^{\text{out}}_L\hat a^{\text{out}}_L(\omega)e^{i\omega t} + \phi^{\text{out}}_{Rp}\hat a^{\text{out}}_{Rp}(\omega)e^{i\omega t} \nonumber\\  \label{eq:decomp2}
    &+ \phi^{\text{out}}_{Rn}(\hat a_{Rn}^{\text{out}}(-\omega))^\dagger e^{-i\omega t}. 
\end{align}
  
\end{subequations}

\noindent Here, $(\hat a^{\text{in}}_i)^{(\dagger)}$ and $(\hat a^{\text{out}}_i)^{(\dagger)}$ are the creation and annihilation operators of the scattering states.
The last term in the decomposition is an excitation of negative energy compared to the other excitations and therefore needs an additional minus sign in front of the frequency for it to hold the same energy as the other terms.
The annihilation operators define a ground state, $\hat a^{\text{in}}_i|0,\text{in}\rangle=0$ and $\hat a^{\text{out}}_i|0,\text{out}\rangle=0$.
We map the in-state basis on the out-state basis through the scattering matrix

\begin{equation}\label{in-out}
    \begin{pmatrix}
        \phi^{\text{out}}_L \\
        \phi^{\text{out}}_{Rp} \\
        \phi^{\text{out}}_{Rn}
    \end{pmatrix}=S\begin{pmatrix}
        \phi^{\text{in}}_L \\
        \phi^{\text{in}}_{Rp} \\
        \phi^{\text{in}}_{Rn}
    \end{pmatrix}.
\end{equation}

\noindent The $S$-matrix is of the form

\begin{align}
    S= & \begin{pmatrix}
        R^*(\omega) & (\Tilde T'(\omega))^* & T^*(\omega) \\
        \Tilde T^*(\omega) & \Tilde R^*(\omega) & 0 \\
        (T'(\omega))^* & 0 & (R'(\omega))^*
    \end{pmatrix}.\label{eq:preS}
\end{align}

\noindent Rather than the conventional unitarity condition, the scattering matrix satisfies the modified unitarity condition, $S\eta S^\dagger =\eta$ with $\eta=\text{diag}(1,1,-1)$.
This is due to the Bogoliubov coefficients assigned to the plane-wave solutions of each in and out-going magnon. 
The relation between Bogoliubov coefficients is conserved and imposes a norm on the scattering coefficients; specifically, we have $|R|^2+|\Tilde T|^2-|T'|^2=1$.
Thus, the $S$-matrix has the $(+1,+1,-1)$ signature.
The $S$-matrix together with the decompositions in Eq. (\ref{eq:decomp1}-\ref{eq:decomp2}) give the relations between the creation and annihilation operators that we associate with each scattering state

\begin{equation} \label{S-matrixOp}
    \begin{pmatrix}
        \hat a^{\text{out}}_L(\omega) \\
        \hat a^{\text{out}}_{Rp}(\omega) \\
        (\hat a^{\text{out}}_{Rn}(-\omega))^\dagger
    \end{pmatrix}=\eta S^*\eta\begin{pmatrix}
        \hat a^{\text{in}}_L(\omega) \\
        \hat a^{\text{in}}_{Rp}(\omega) \\
        (\hat a^{\text{in}}_{Rn}(-\omega))^\dagger
    \end{pmatrix}.
\end{equation}

\noindent This result provides the relation that is used in Sec. \ref{sec:General Current expression} to compute the current. 
As we shall see below, explicit computation shows that this scattering matrix simplifies further. 
The $S$-matrix mixes creation and annihilation operators, which is characteristic of Hawking radiation \cite{Hawking1975} and is expanded upon in Appendix \ref{sec:Magnonnumber}.

\begin{figure}[t]
    \centering
    \tikzset{every picture/.style={line width=0.75pt}} 

\begin{tikzpicture}[x=0.65pt,y=0.65pt,yscale=-1,xscale=1]

\draw [color={rgb, 255:red, 218; green, 214; blue, 74 }  ,draw opacity=1 ] [dash pattern={on 4.5pt off 4.5pt}]  (331.15,68.73) -- (330.75,219.53) ;
\draw [color={rgb, 255:red, 74; green, 144; blue, 226 }  ,draw opacity=1 ]   (247.15,186.33) -- (326.26,135.83) ;
\draw [shift={(327.95,134.75)}, rotate = 147.45] [color={rgb, 255:red, 74; green, 144; blue, 226 }  ,draw opacity=1 ][line width=0.75]    (10.93,-3.29) .. controls (6.95,-1.4) and (3.31,-0.3) .. (0,0) .. controls (3.31,0.3) and (6.95,1.4) .. (10.93,3.29)   ;
\draw [color={rgb, 255:red, 74; green, 144; blue, 226 }  ,draw opacity=1 ]   (328.75,131.05) -- (251.99,77.87) ;
\draw [shift={(250.35,76.73)}, rotate = 34.72] [color={rgb, 255:red, 74; green, 144; blue, 226 }  ,draw opacity=1 ][line width=0.75]    (10.93,-3.29) .. controls (6.95,-1.4) and (3.31,-0.3) .. (0,0) .. controls (3.31,0.3) and (6.95,1.4) .. (10.93,3.29)   ;
\draw [color={rgb, 255:red, 74; green, 144; blue, 226 }  ,draw opacity=1 ]   (335.15,134.75) -- (408.7,185.2) ;
\draw [shift={(410.35,186.33)}, rotate = 214.44] [color={rgb, 255:red, 74; green, 144; blue, 226 }  ,draw opacity=1 ][line width=0.75]    (10.93,-3.29) .. controls (6.95,-1.4) and (3.31,-0.3) .. (0,0) .. controls (3.31,0.3) and (6.95,1.4) .. (10.93,3.29)   ;
\draw [color={rgb, 255:red, 208; green, 2; blue, 27 }  ,draw opacity=1 ]   (334.75,130.65) -- (406.11,77.52) ;
\draw [shift={(407.71,76.33)}, rotate = 143.33] [color={rgb, 255:red, 208; green, 2; blue, 27 }  ,draw opacity=1 ][line width=0.75]    (10.93,-3.29) .. controls (6.95,-1.4) and (3.31,-0.3) .. (0,0) .. controls (3.31,0.3) and (6.95,1.4) .. (10.93,3.29)   ;
\draw [line width=0.75]    (275.95,211.53) .. controls (311.55,212.73) and (322.35,203.53) .. (327.55,193.13) ;

\draw (230.8,79.2) node [anchor=north west][inner sep=0.75pt]  [font=\footnotesize,color={rgb, 255:red, 74; green, 144; blue, 226 }  ,opacity=1 ]  {$A_{l}^{L}$};
\draw (210.2,165.2) node [anchor=north west][inner sep=0.75pt]  [font=\footnotesize,color={rgb, 255:red, 74; green, 144; blue, 226 }  ,opacity=1 ]  {$A_{r}^{L}=1$};
\draw (414.4,165.2) node [anchor=north west][inner sep=0.75pt]  [font=\footnotesize,color={rgb, 255:red, 74; green, 144; blue, 226 }  ,opacity=1 ]  {$A_{r}^{R}$};
\draw (409.71,79.73) node [anchor=north west][inner sep=0.75pt]  [font=\footnotesize,color={rgb, 255:red, 208; green, 2; blue, 27 }  ,opacity=1 ]  {$A_{+}^{R}$};
\draw (254.8,204) node [anchor=north west][inner sep=0.75pt]  [font=\footnotesize,color={rgb, 255:red, 0; green, 0; blue, 0 }  ,opacity=1 ]  {$A_{-}^{L}$};

\end{tikzpicture}
    \caption{Scattering process with one incoming magnon. $A^{L/R}_i$ are the amplitudes of the modes with superscripts denoting left/right FM and subscripts denoting their momentum modes. Adapted from \cite{Faccio:2013kpa}.}
    \label{fig:onemagnonscattering}
\end{figure}
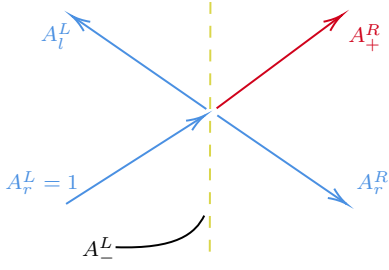

In the remainder of this section, we compute the entries of the scattering matrix, which are the reflection and transmission coefficients of the scattering states.
We substitute Eq. (\ref{eq:decomposition}) into Eq. (\ref{eq:BEOM1}-\ref{eq:BEOM2}) to find

\begin{subequations}
\begin{align}
    (\partial_{x}-\gamma) \phi_L(\omega,0)&=\gamma  \varphi_R(\omega,0), \label{BEOMB1}\\
    (\partial_{x}-\gamma) \varphi_L(\omega,0)&=\gamma \phi_R(\omega,0), \\
    -(\partial_{x}+\gamma) \phi_R(\omega,0)&=\gamma \varphi_L(\omega,0), \\
    -(\partial_{x}+\gamma) \varphi_R(\omega,0)&=\gamma \phi_L(\omega,0). \label{BEOMB2}
\end{align}
\end{subequations}

\noindent Subsequently, we substitute the plane wave decomposition of Eq. (\ref{aLi1}-\ref{aLi2}) into the conditions above and find the matrix equation

\begin{equation}
    W_L\begin{pmatrix}
    A_r^L \\
    A_l^L \\
    A_+^L \\
    A_-^L \\
    \end{pmatrix}=
        W_R\begin{pmatrix}
    A_r^R \\
    A_l^R \\
    A_+^R \\
    A_-^R \\
    \end{pmatrix},\label{ampl}
\end{equation}

\noindent with matrices $W_L$ and $W_R$ equal to

\begin{subequations}
    
\begin{align}
    W_L&=\begin{pmatrix}
    \lambda^L_r & \lambda^L_l & 0 & 0 \\
    \gamma & \gamma & 0 & 0 \\
    0 & 0 & \lambda^L_+ & \lambda^L_- \\
    0 & 0 & \gamma & \gamma \\
    \end{pmatrix}, \\
    W_R&=\begin{pmatrix}
    0 & 0 & \gamma & \gamma \\
    0 & 0 & \lambda^R_+ & \lambda^R_- \\
    \gamma & \gamma & 0 & 0 \\
    \lambda^R_r & \lambda^R_l & 0 & 0 \\
    \end{pmatrix}.
\end{align}

\end{subequations}

\noindent Here $\lambda^L_i=ik_i^L-\gamma$ and $\lambda^R_i=-ik_i^R-\gamma$.
This allows us to compute the amplitudes of any combination of magnon modes.

For example, we compute one set of amplitudes corresponding to the scattering state in Fig. \ref{fig:onemagnonscattering}.
There is an incoming magnon from the left FM with amplitude $A_r^L=1$.
Then, we have four possible outgoing magnons with amplitudes $A_l^L, A_r^R, A_+^R$ and $A_-^L$.
This reduces Eq. (\ref{ampl}) to

\begin{equation}
    W_L\begin{pmatrix}
    1 \\
    A_l^L \\
    0 \\
    A_-^L \\
    \end{pmatrix}=
       W_R\begin{pmatrix}
    A_r^R \\
    0 \\
    A_+^R \\
    0 \\
    \end{pmatrix},
\end{equation}

\noindent with the solution

\begin{subequations}

\begin{align}
    A_-^L=&0, \\
    A_r^R=&0, \\
    A_l^L=&-\frac{\gamma^2-\lambda^R_+\lambda^L_r}{\gamma^2-\lambda^R_+\lambda^L_l},\\
    A_+^R=&\frac{1}{\lambda^R_+-\lambda^R_-}\bigg[(\gamma^2-\lambda^R_-\lambda^L_r) \nonumber\\
    &-\frac{\gamma^2-\lambda^R_+\lambda^L_r}{\gamma^2-\lambda^R_+\lambda^L_l}(\gamma^2-\lambda^R_-\lambda^L_l)\bigg]. \label{notT}
\end{align}
\end{subequations}

\noindent The reflection and transmission coefficients are $|R(\omega)|^2= |A^L_l|^2$ and $|T(\omega)|^2=|A^R_+|^2|k^R_+/k^L_r|$.
The scattering coefficients $|R(\omega)|^2$ and $|T(\omega)|^2$ are therefore given by

\begin{widetext}
    \begin{equation}
    |R(\omega)|^2=\frac{\gamma^2 \left(h_R-h_L\right)+(h_R-\omega) (\omega-h_L)+2\gamma^2 \sqrt{\omega-h_L} \sqrt{h_R-\omega}}{\gamma^2 \left(h_R-h_L\right)+(h_R-\omega) (\omega-h_L)-2\gamma^2 \sqrt{\omega-h_L} \sqrt{h_R-\omega}},
\end{equation}
\begin{equation}
    |T(\omega)|^2=\frac{4 \gamma^2 \sqrt{\omega-h_L} \sqrt{h_R-\omega}}{\gamma^2(h_R-h_L)+(h_R-\omega)(\omega-h_L)-2 \gamma^2 \sqrt{\omega-h_L} \sqrt{h_R-\omega}}.
\end{equation}
\end{widetext}

\noindent These expressions are only appropriate for $h_L<\omega<h_R$, which guarantees that the term $2\gamma^2 \sqrt{\omega-h_L} \sqrt{h_R-\omega}$ is real and positive.
This implies enhanced reflection, $|R(\omega)|^2>1$.
This remarkable property was already discussed in Ref. \cite{PhysRevApplied.18.064026}.

We compute all other reflection and transmission coefficients, similar to the example of Fig. \ref{fig:onemagnonscattering}.
This results in the relations $|T(\omega)|^2=|T'(\omega)|^2$, $|R(\omega)|^2=|R'(\omega)|^2$ and $|\Tilde R(\omega)|^2=1$, $|\Tilde T(\omega)|^2=|\Tilde T'(\omega)|^2=0$, for $h_L<\omega<h_R$.
Therefore the $S$-matrix simplifies to

\begin{align}
    S= & \begin{pmatrix}
        R^*(\omega) & 0 & T^*(\omega) \\
        0 & 1 & 0 \\
        (T'(\omega))^* & 0 & (R'(\omega))^*
    \end{pmatrix}.\label{eq:S}
\end{align}

\noindent In this $S$-matrix the positive energy excitations in the right FM are decoupled and therefore it reduces to the scattering matrix given in Eq. (\ref{eq:GenSmatrix}), with $R_L(\omega)=R(\omega)$, $T_L(\omega)=-T(\omega)$, $T_R(\omega)=-T'(\omega)$ and $R_R(\omega)=R'(\omega)$.

\section{Spontaneous magnon pair production}\label{sec:Spontaneous emission}

In Section \ref{sec:General Current expression}, we generalized the Landauer-B\"uttiker formalism to a scattering matrix that mixes creation and annihilation operators as is proven to be the case for the set-up in Fig. \ref{fig:setup}.
In this section, we use this result to compute the spin currents and their correlation.

We treat the left reservoir of the driven synthetic antiferromagnet as a spin sink with zero spin accumulation. 
The right lead has spin accumulation $-\Delta \mu>h_R$ to dynamically stabilize the right FM \cite{antimagnonics}. 
We assume both leads have equal temperature $T=1/k_B \beta$ with $k_B$ the Boltzmann constant. 
The distribution functions are then given by  

\begin{subequations}

\begin{align} \label{fL}
    f_{L}(\omega,\beta)&=\frac{1}{e^{\beta\omega}-1}, \\
    f_{R}(\omega,\beta)&=\frac{1}{e^{\beta(\omega+\Delta\mu)}-1}.
\end{align}

\end{subequations}

\noindent We only consider frequencies for which the negative and positive energy magnons can couple, $h_L<\omega<h_R$.
The magnon currents are computed through Eq. (\ref{LBform}) which yields

\begin{subequations}
    
\begin{align}
    \langle\hat J_{L}\rangle&=-\int d\omega |T(\omega)|^2(1+f_{L}(\omega,\beta)+f_{R}(-\omega,\beta)),\label{current} \\
    \langle\hat J_{R}\rangle&=-\int d\omega|T(\omega)|^2(1+f_{R}(-\omega,\beta)+f_{L}(\omega,\beta)).\label{anticurrent}
\end{align}
\end{subequations}

\noindent Notice that there is an overall minus sign, which implies that all magnon currents are flowing into the reservoirs.
The magnon currents have a term that is independent of the distributions.
Thus, even when there are no thermally excited magnons, we expect spontaneously excited magnons from the interface.
This effect is the same in both FMs, which implies that the magnons are generated in pairs.
We expect that the correlation shows entanglement.
We define the current-current correlation as

\begin{equation}
    C_{\alpha\beta}(\tau)\equiv \langle \hat J_{\alpha}(t)\hat J_{\beta}(t+\tau)\rangle -\langle\hat{J}_\alpha(t)\rangle\langle\hat{J}_\beta(t+\tau)\rangle.
\end{equation}

\noindent Thus the correlation between the left and right spin current is

\begin{align} 
     &C_{LR}(\tau)=\int d\omega d\omega' \big(-T^*(\omega)T(\omega')(1- R(\omega)R^*(\omega'))\nonumber \\
     &\times[f_L(\omega)(1+f_L(\omega'))+f_{R}(-\omega')(1+f_{R}(-\omega))] \nonumber \\
     &+T^*(\omega)T(\omega')R(\omega)R^*(\omega') \nonumber \\
     &\times [1+f_L(\omega')(1+f_{R}(-\omega'))+f_{R}(-\omega)+f_L(\omega)f_{R}(-\omega')]\big) \nonumber \\
     &\times e^{-i (\omega-\omega') \tau}.\label{eq:currentcorr}
\end{align}

\noindent Consider the case of vanishing distributions, then the correlation is reduced to $\int T^*(\omega)T(\omega')R(\omega)R^*(\omega')e^{-i (\omega-\omega') \tau}d\omega d\omega'$.
This represents the correlations due to magnon pairs generated at the interface.
In the next section, we show that this correlation violates classical inequalities.
In Appendix \ref{sec:powerspectrum}, we compute the zero-frequency power spectrum that is associated with the current-current correlation.

\subsection{Violation of Cauchy-Schwarz Inequalities}

We demonstrate the entanglement of magnon pairs by proving that their correlation violates the Cauchy-Schwarz (CS) inequality.
In quantum optics \cite{QO:loudon_quantum_2000,QO:walls_quantum_2008} it is shown that correlation functions of classical light satisfy a CS inequality.
However, a strongly non-classical correlation can violate the inequality.
In particular, de Nova \textit{et. al.} demonstrated that the violation of the Cauchy-Schwarz (CS) inequality indicates the non-separability of two states, provided that these states are both Gaussian and incoherent \cite{Nova_2015}, which is the situation that is relevant to us.
In Appendix \ref{sec:CS-inequality}, we prove that the CS inequality is violated if the following inequality is also violated

\begin{equation}\label{eq:CS-corr}
    C_{LR}(0)-\langle J_L(0)\rangle_0\langle J_{Rn}(0)\rangle_0\leq 0.
\end{equation}

\noindent The expectation values are computed in the vacuum in-state ($a_i^{\text{in}}|0,\text{in}\rangle=0$).
This inequality is violated for the results in Eq. (\ref{current}-\ref{anticurrent}) and Eq. (\ref{eq:currentcorr}).
Thus the magnon pairs are non-separable and therefore entangled.

\section{Discussion and Conclusion} \label{sec:Discussion and Conclusion}

In this article, we analytically showed that a driven synthetic antiferromagnet subject to an inhomogeneous field as in Fig. \ref{fig:setup} emits entangled magnon pairs.
This phenomenon arises because of coupling between positive-energy and negative-energy magnons. 
The latter exist in magnets that are dynamically stabilized with their magnetization direction against the effective field.
Therefore, the analysis applies to a broad range of materials, provided that driving can stabilize spins opposite to the magnetic field.
Notably, effects like DMI and anisotropy were not taken into account. 
Nevertheless, it is worth noting that anisotropy affects our results only qualitatively. 
Additionally, in Ref. \cite{PhysRevApplied.18.064026}, numerical simulations demonstrated only a slight alteration in the results when DMI was considered at the interface. 

All our analysis is done in the lowest-order approximation of the Holstein-Primakoff transformation.
Thus, interactions such as the magnon-magnon interaction or magnon-phonon interactions are not considered.
Potentially, this could enhance or lower the effect.
However, for low temperatures, the analysis in this article will be sufficient.

To observe the magnon entanglement, the system has to be cooled sufficiently such that the quantum effects are not overwhelmed by thermal fluctuations. 
To estimate the required temperatures, consider the magnon current in Eq. (\ref{current}).
It has two contributions, the generation of magnon pairs and thermal fluctuations of left and right magnons.
The crossover temperature is defined as the temperature for which both contributions to the current are equal, i.e.,

\begin{equation}
    |T(\omega)|^2=|T(\omega)|^2f_{L}(\omega,\beta_c)+|T(\omega)|^2f_{R}(-\omega,\beta_c).
\end{equation}

\noindent Here, $\beta_c=1/k_B T_c$ with $T_c$ the crossover temperature.
The generated magnons exist in the frequency window $h_L<\omega<h_R$.
For typical magnetic fields, $h_L$ and $h_R$ are on the order of gigahertz which yields temperatures in the millikelvin range. 
For larger fields, the temperature below which the predicted magnon entanglement occurs will increase. 
Similarly, if we replace the FMs with antiferromagnets, which allow for larger frequencies, then the crossover temperature will increase.
This setup has been investigated classically by Ref. \cite{AFM:errani_negative-energy_2024,AFM:yuan_magnonic_2023}.

Upon injecting angular momentum in the right FM using the spin Hall effect, it will inevitably heat up.
However, the magnon pairs are also generated without driving in the transient regime. 
Starting without a magnetic field, the synthetic antiferromagnet is in its antiparallel ground state. 
Upon switching on the inhomogeneous magnetic field, the entangled magnon pair production starts before the right ferromagnet reverses to point along the effective field. 
For small temperatures $T \ll T_c$ this classical magnetization reversal will take much longer than the pair production. 
This should make it possible to observe the entangled magnons without Joule heating the system by the current that drives the spin-orbit torque.

In future investigations, we aim to explore the relationship between entanglement and anisotropy, given its role in magnon squeezing \cite{QM:Kamra}.
This could increase the entanglement between the pairs, similar to what Brady \textit{et. al}. have shown for cold-atom systems \cite{Brady}.
Improving the entanglement of this set-up would increase its applicability in future technologies such as quantum computing and quantum communication.
A recent work, for example, considers the entanglement of NV centers in a set-up that is similar to the one that we propose in this article \cite{kleinherber}. 

\begin{acknowledgments}
This work is funded by the projects “Black holes on a chip” with project number OCENW.KLEIN.502 and “Fluid Spintronics” with project number VI.C.182.069.
Both are financed by the Dutch Research Council (NWO).
\end{acknowledgments}

\appendix

\section{Current calculations}\label{currentApp}

In this appendix, we compute the spin current in Sec. \ref{sec:General Current expression}.
However, we consider a more general set-up.
This set-up generalizes the synthetic antiferromagnet considered previously to have $n$ leads.
Within each lead, we consider the presence of two magnon bands: one with positive-energy excitations and one with negative-energy ones. 
The operators associated with these bands undergo a mixing process.
Define the in and out-state in each lead as $\phi^{\text{in}}_{\alpha i}(\omega,x)$ and $\phi^{\text{out}}_{\alpha i}(\omega,x)$, where $\alpha$ gives the lead and $i$ gives the band.
In the simplified set-up, $\alpha$ denoted the left or right lead of the synthetic antiferromagnet, $\alpha\in\{L,R\}$.
The operator $(\hat a^{\text{in}}_{\alpha i}(\omega))^\dagger$ excites $\phi^{\text{in}}_{\alpha i}/\sqrt{\hbar v_{\alpha i}}$ and $(\hat a^{\text{out}}_{\alpha i}(\omega))^\dagger$ excites $\phi^{\text{out}}_{\alpha i}/\sqrt{\hbar v_{\alpha i}}$ with $v_{\alpha i}=\hbar k_{\alpha i}/m$ which normalizes the flux.
The operators obey the bosonic commutation relations.
Therefore, the general scattering matrix is

\begin{equation}
    \begin{pmatrix}
        \hat a^{\text{out}}_{11} \\
        \vdots \\
        \hat a^{\text{out}}_{n1} \\
        (\hat a^{\text{out}}_{12})^\dagger \\
        \vdots \\
        (\hat a^{\text{out}}_{n2})^\dagger
    \end{pmatrix}=
    \begin{pmatrix}
        S_{(\alpha1)(\beta1)} & S_{(\alpha1)(\beta2)} \\
        S_{(\alpha2)(\beta1)} & S_{(\alpha2)(\beta2)} \\
    \end{pmatrix}
    \begin{pmatrix}
        \hat a^{\text{in}}_{11} \\
        \vdots \\
        \hat a^{\text{in}}_{n1} \\
        (\hat a^{\text{in}}_{12})^\dagger \\
        \vdots \\
        (\hat a^{\text{in}}_{n2})^\dagger
    \end{pmatrix}.
\end{equation}

\noindent Where $\alpha,\beta=1\dots n$ and span the matrices $S_{(\alpha1)(\beta1)}, S_{(\alpha1)(\beta2)}, S_{(\alpha2)(\beta1)}$ and $S_{(\alpha2)(\beta2)}$.
$S_{(\alpha1)(\beta1)}$ and $S_{(\alpha2)(\beta2)}$ are the scattering matrices between the same band.
$S_{(\alpha1)(\beta2)}$ and $S_{(\alpha2)(\beta1)}$ are the matrices that mix bands.
The diagonal component of the scattering matrix corresponds to reflection of the incoming wave, $S^\dagger_{(\alpha i)(\alpha j)}S_{(\alpha i)(\alpha j)}=|R_{\alpha i}(\omega)|^2\delta_{ij}$.
The other entries are transmission coefficients.
In order to compute the current, we need to define field operators.

\begin{align}
    \hat \Phi_{\alpha i}(\omega,x)&=\hat a^{\text{in}}_{\alpha i}(\omega)\frac{\phi^{\text{in}}_{\alpha i}(\omega,x)}{\sqrt{\hbar v_{\alpha i}}}+\hat a^{\text{out}}_{\alpha i}(\omega)\frac{\phi^{\text{out}}_{\alpha i}(\omega,x)}{\sqrt{\hbar v_{\alpha i}}}, \\
    \hat \Phi^\dagger_{\alpha i}(\omega,x)&=(\hat a^{\text{in}}_{\alpha i}(\omega))^\dagger\frac{\phi^{\text{in}*}_{\alpha i}(\omega,x)}{\sqrt{\hbar v_{\alpha i}}}+(\hat a^{\text{in}}_{\alpha i}(\omega))^\dagger\frac{\phi^{\text{out}*}_{\alpha i}(\omega,x)}{\sqrt{\hbar v_{\alpha i}}}.
\end{align}

\noindent The current is defined as

\begin{align}\label{eq:currentEApp}
    \hat J_{\alpha i}(t)=&\frac{JSa^2}{i\hbar}\int d\omega d\omega' (\hat\Phi_{\alpha i}^\dagger(\omega,x)\partial_x\hat\Phi_{\alpha i}(\omega',x) \nonumber\\
    &-(\partial_x\hat\Phi_{\alpha i}^\dagger(\omega,x))\hat\Phi_{\alpha i}(\omega',x))e^{i(\omega-\omega')t/\hbar}, \nonumber\\
    =&\frac{1}{\hbar}\int d\omega d\omega'((\hat a^{\text{in}}_{\alpha i}(\omega))^\dagger\hat a^{\text{in}}_{\alpha i}(\omega') \nonumber\\
    &-(\hat a^{\text{out}}_{\alpha i}(\omega))^\dagger\hat a^{\text{out}}_{\alpha i}(\omega')) e^{i(\omega-\omega')t/\hbar}. 
\end{align}

\noindent We have used that the in and out-states have opposite momentum. 
Also, we neglected the energy dependence of the velocities, since they vary slowly with energy, as is explained on pages 11-12 of Ref. \cite{BLANTER20001}.
The quantum average of the number operator is $\langle\hat a_{\alpha i}^\dagger(\omega)\hat a_{\beta j}(\omega')\rangle= \delta_{\alpha\beta}\delta_{ij}\delta(\omega-\omega') f_{\alpha i}(\omega)$ where $f_{\alpha i}(\omega)$ is the distribution.
Form Eq. (\ref{eq:currentEApp}) we find that

\begin{align}
    \langle\hat J_{\alpha i}\rangle=& \frac{1}{\hbar}\int d\omega\bigg(\sum_{\beta\neq\alpha}[|T_{(\alpha i)(\beta i)}(\omega)|^2(f_{\alpha i}(\omega)-f_{\beta i}(\omega))] \nonumber \\
    &-|T_{(\alpha i) (\beta j)}(\omega)|^2(1+f_{\alpha i}(\omega)+f_{\beta j}(\omega))\bigg). \label{LBformAp}
\end{align}

\noindent where the condition $i\neq j$ ensures that the entries are from different bands.
The first term is the typical Landauer-B\"uttiker formula, and the second term is due to the mixing of creation and annihilation operators.
This term persists even when the distributions of magnons vanish.
This alludes to the spontaneous excitation of magnons on the interface.

\section{Synthetic ferromagnet}\label{noSOT}

In this appendix, we consider the original set-up without driving.
This causes the spins to find the energetically favourable configuration, which is all spins pointing along the external magnetic field.
Therefore, the new setup is a synthetic ferromagnet in an inhomogeneous magnetic field.
We have the Hamiltonian

\begin{align}
    \hat H=&-J\sum_{\langle i,j\rangle\in (U_L \vee U_R)} \Vec{S}_i\cdot\Vec{S}_j  \nonumber \\
    &- h_L\sum_{i\in U_L} S^z_i - h_R\sum_{i\in U_R}S^z_i.
\end{align}

\noindent Here, $U_{L/R}$ is the set of lattice positions on the left and right sides, respectively.
We apply the Holstein-Primakoff transformation and linearize to quadratic order and take the continuous limit.
The Hamiltonian takes the following form

\begin{align}
    \hat H=&\sum_{\nu\in\{L,R\}}\frac{\hbar }{2}\int dx \hspace{3px}\psid_\nu(x,t)(-JSa^2\partial_x^2+ h_\nu)\hat \psi_\nu(x,t) \nonumber\\
    &+h.c.,
\end{align}

\noindent where $\hat \psi^{(\dagger)}_{R/L}(x,t)$ is a continuous magnon annihilation (creation) operator.
The next step is to substitute the Fourier transformation which gives us the dispersion relation in the left FM

\begin{equation}
    \omega^L_{k}=JSa^2k^2+h_{L}.
\end{equation}

\noindent Similarly for the right FM

\begin{equation}
    \omega^R_{k}=JSa^2k^2+h_{R}.
\end{equation}

\noindent The wave vectors of magnons in the left FM have two real solutions and two imaginary solutions for $\omega>h_L$

\begin{align}
    \Lambda k^L_{r/l}&=\pm\sqrt{\omega -h_L}, & \Lambda k^L_{+/-}&=\pm i\sqrt{\omega +h_L}.
\end{align}

\noindent Here $\Lambda^2={JS}a^2$ and $k^L_{r/l}$ describe the right ($+\sqrt{\omega -h_L}$) and left ($-\sqrt{\omega -h_L}$) moving magnon mode.
$k^L_{+/-}$ describes the exponentially damping and growing mode. 
The exponentially growing mode is non-physical and is therefore not included as a possible scattering outcome.
For the right FM, we have two real solutions and two imaginary solutions for $\omega>h_R$

\begin{align}
    \Lambda k^R_{r/l}&=\pm\sqrt{\omega -h_R}, & \Lambda k^R_{+/-}&=\pm i\sqrt{\omega +h_R}.
\end{align}

\noindent $k^R_{r/l}$ describe the right ($+\sqrt{\omega -h_R}$) and left ($-\sqrt{\omega -h_R}$) moving magnon mode.
$k^R_{+/-}$ describes the exponentially damping and growing mode.

\subsection{Boundary conditions}

In this section, we start with including a boundary term in the Hamiltonian.
The boundary term completes the Hamiltonian,

\begin{align}
    \Ha=&\Ha_L+\Ha_R+J'\Vec{S}_{L}(0)\cdot\Vec{S}_{R}(0).
\end{align}

\noindent We previously showed the dispersion relations in the bulk.
The next step is to compute the conditions imposed by the boundary term.
We start by using the Holstein-Primakoff transformation to find a boundary term for magnon operators.
Then we compute the Heisenberg equation of motion for the continuous magnon operator.
For the magnon operator in the left FM we find

\begin{align}
    \partial_{x_B}\hat \psi_L(x_B,t)-\gamma(\hat \psi_L(x_B,t)+\hat \psi_R(x_B,t))&=0
\end{align}

\noindent where $\gamma=J'/2Ja$.
We do this treatment again for the right FM and find

\begin{align}
    \partial_{x_B}\hat \psi_R(x_B,t)+\gamma(\hat \psi_R(x_B,t)+\hat \psi_L(x_B,t))&=0.
\end{align}

\subsection{$S$-matrix}

As in main part of the paper, we need to decompose the operators in positive and negative frequencies

\begin{align}
    \hat \psi_\nu(x,t) &=\int d\omega \hspace{3px}  \hat a(\omega) \phi_\nu(\omega,x) e^{i\omega t}+\hat a^\dagger(\omega) (\varphi_{\nu}(\omega,x))^* e^{-i\omega t}
\end{align}

\noindent Here $\nu\in\{L,R\}$. This decomposition leads to the boundary conditions

\begin{align}
    (\partial_{x}-\gamma)\phi_L(\omega,0)&=\gamma\phi_R(\omega,0), \\
    (\partial_{x}-\gamma)\varphi_L(\omega,0)&=\gamma\varphi_R(\omega,0), \\
    -(\partial_{x}+\gamma)\phi_R(\omega,0)&=\gamma \phi_L(\omega,0), \\
    -(\partial_{x}+\gamma)\varphi_R(\omega,0)&=\gamma\varphi_L(\omega,0). 
\end{align}

\noindent The positive and negative frequency solutions do not couple in contrast with the set-up considered in the paper.
They are expanded into momentum modes

\begin{align} 
    \phi_{\nu}(\omega,x)&=\sum_i A_{i}^{\nu}u_{i}^{\nu}e^{ik^{\nu}_i(\omega)x}, \\ 
    \varphi_{\nu}(\omega,x)&=\sum_i A_{i}^{\nu}v_{i}^{\nu}e^{ik^{\nu}_i(\omega)x}.
\end{align} 

\noindent $A_{i}^{\nu}$ are the amplitudes of each magnon mode where $i$ sums over the two possible wave number solutions for a given $\omega$, $i\in\{l,r\}$.
$u_{i}^{\nu}$ and $v_{i}^{\nu}$ are Bogoliubov modes that satisfy $|u_{i}^{\nu}|^2-|v_{i}^{\nu}|^2=1$.
$\hat a^{(\dagger)}_{\nu}(\omega)$ is the annihilation (creation) operator of the magnon mode and is a bosonic operator.
We solve for the bulk equations of motion and we find the following conditions

\begin{align}
    u^{\nu}_{l/r}&=1, & u^{\nu}_{+/-}&=0, \\
    v^{\nu}_{l/r}&=0, & v^{\nu}_{+/-}&=1.
\end{align}

\noindent Now we are in the position to reduce the boundary conditions to

\begin{equation}
    \begin{pmatrix}
    \lambda^L_r & \lambda^L_l & 0 & 0 \\
    \gamma & \gamma & 0 & 0 \\
    0 & 0 & \lambda^L_+ & \lambda^L_- \\
    0 & 0 & \gamma & \gamma \\
    \end{pmatrix}
    \begin{pmatrix}
    A_r^L \\
    A_l^L \\
    A_+^L \\
    A_-^L \\
    \end{pmatrix}=
    \begin{pmatrix}
    \gamma & \gamma & 0 & 0 \\
    \lambda^R_r & \lambda^R_l & 0 & 0 \\
    0 & 0 & \gamma & \gamma \\
    0 & 0 & \lambda^R_+ & \lambda^R_- \\
    \end{pmatrix}
    \begin{pmatrix}
    A_r^R \\
    A_l^R \\
    A_+^R \\
    A_-^R \\
    \end{pmatrix}.
\end{equation}

\noindent Here, $\lambda^L_i=ik_i^L-\gamma$ and $\lambda^R_i=-ik_i^R-\gamma$.
We consider a set of scattering wavefunctions that form a complete basis.

\begin{align}
    \phi^{\text{in}}_L(x) &= 
\begin{cases}
    e^{ik^L_rx}+Re^{ik^L_lx}, & \text{if } x<0\\
    Te^{ik^R_rx},              & \text{otherwise.}
\end{cases} \\
    \phi^{\text{in}}_R(x) &= 
\begin{cases}
    T'e^{ik^L_lx},                & \text{if } x<0\\
    e^{ik^R_lx}+R'e^{ik^R_rx},    & \text{otherwise.}
\end{cases}\\
    \phi^{\text{out}}_L(x) &= 
\begin{cases}
    e^{ik^L_lx}+R^*e^{ik^L_rx}, & \text{if } x<0\\
    T^*e^{ik^R_lx},              & \text{otherwise.}
\end{cases} \\
    \phi^{\text{out}}_R(x) &= 
\begin{cases}
    T'^*e^{ik^L_rx},                & \text{if } x<0\\
    e^{ik^R_rx}+R'^*e^{ik^R_lx},    & \text{otherwise.}
\end{cases}
\end{align}

\noindent The in/out index says if the scattering describes a single in/outgoing magnon with the associated scattered magnons.
These relate to each other via the scattering matrix

\begin{equation}
    \begin{pmatrix}
        \phi^{\text{out}}_L \\
        \phi^{\text{out}}_{R}
    \end{pmatrix}=S^*\begin{pmatrix}
        \phi^{\text{in}}_L \\
        \phi^{\text{in}}_{R}
    \end{pmatrix}.
\end{equation}

\noindent The scattering matrix is defined as

\begin{align}
    S= & \begin{pmatrix}
        R & T \\
        T' & R'
    \end{pmatrix}.
\end{align}

\noindent Define the ingoing scattered state creation and annihilation operator with $(\hat a_{\nu}^{\text{in}})^\dagger$ and $\hat a^{\text{in}}_{\nu}$, respectively.
We do the same for the outgoing scattered state $(\hat a_{\nu}^{\text{out}})^\dagger$ and $\hat a^{\text{out}}_{\nu}$, respectively.
These are related to each other in the same manner as the scattering states.

\begin{equation}\label{S-matrixOpnoSOT}
    \begin{pmatrix}
        \hat a^{\text{out}}_{L} \\
        \hat a^{\text{out}}_{R}
    \end{pmatrix}=S^*\begin{pmatrix}
        \hat a^{\text{in}}_{L} \\
        \hat a^{\text{in}}_{R}
    \end{pmatrix}.
\end{equation}

\subsection{Average current}

In this appendix, we compute the expectation value of the current in the synthetic antiferromagnet without driving.
Eq. (\ref{eq:currentEApp}) allows us to compute the expectation value of the current in each lead we find

\begin{align}
    \langle\hat J_{\alpha}\rangle= \frac{1}{\hbar}\int dE\sum_{\beta\neq \alpha}[|T_{(\alpha)(\beta)}(\omega)|^2(f_{\alpha}(\omega)-f_{\beta}(\omega))] .
\end{align}

\noindent Here $f_\alpha$ is the Bose-Einstein distributions.
This result is in accordance with the usual Landauer-B\"uttiker formalism.

\section{Power spectrum}\label{sec:powerspectrum}

The power spectrum is commonly used in signal processing techniques.
It can give valuable information about the content of a signal, which includes noise.
By examining the power spectrum, we can analyze the frequencies that contribute to the noise of the signal.
The power spectrum is the Fourier transform of the current-current correlation function \cite{BLANTER20001}

\begin{equation}
    P(\omega)=\int d\tau C_{LRn}(\tau) e^{-i \omega \tau}.
\end{equation}

\noindent We are mainly interested in the zero-frequency power spectrum since it provides insight into the noise on long-time scales

\begin{align}
    P(0)&=\int d\omega \big(|T(\omega)|^4 [(1+f_L)f_L+(1+f_{R})f_{R}] \nonumber\\
     &+|T(\omega)|^2|R(\omega)|^2 [1+(1+f_{R})f_L+(1+f_L)f_{R}]\big).
\end{align}

\noindent There are two distinct parts of the zero-frequency power spectrum.
The terms that depend on the distributions are associated with thermal noise.
There is also a term that is independent of temperature, $\int d\omega|T(\omega)|^2|R(\omega)|^2$.
This is the noise associated with the pair production on the interface.

\section{CS-inequality}\label{sec:CS-inequality}

De Nova \textit{et. al.} defined the CS-inequality to be

\begin{align}\label{eq:CSinequality}
    &\langle (\hat a_L^{\text{out}})^\dagger (\hat a_{Rn}^{\text{out}})^\dagger\hat a_L^{\text{out}}\hat a_{Rn}^{\text{out}}\rangle_0^2 \\
    &-\langle (\hat a_L^{\text{out}})^\dagger(\hat a_L^{\text{out}})^\dagger\hat a_L^{\text{out}}\hat a_L^{\text{out}}\rangle_0\langle (\hat a_{Rn}^{\text{out}})^\dagger (\hat a_{Rn}^{\text{out}})^\dagger\hat a_{Rn}^{\text{out}}\hat a_{Rn}^{\text{out}}\rangle_0\leq 0. \nonumber
\end{align}

\noindent The expectation values are computed with the vacuum in-state ($a_i^{\text{in}}|0,\text{in}\rangle=0$).
In this section, we prove that this inequality is violated if Eq. (\ref{eq:CS-corr}) is violated.
We use that 

\begin{align}\label{eq:term1}
    &(C_{LRn}(0)+\langle J_L(0)\rangle_0\langle J_{Rn}(0)\rangle_0)^2 \\
    &=\int d\omega d\omega'\langle (\hat a_L^{\text{out}}(\omega))^\dagger (\hat a_{Rn}^{\text{out}}(\omega'))^\dagger\hat a_L^{\text{out}}(\omega)\hat a_{Rn}^{\text{out}}(\omega')\rangle_0^2. \nonumber
\end{align} 

\noindent We also use that

\begin{align}\label{eq:term2}
    &2\langle J_{L/Rn}(0)\rangle_0^2 \\
    &=\int d\omega d\omega'\langle (\hat a_{L/Rn}^{\text{out}}(\omega))^\dagger(\hat a_{L/Rn}^{\text{out}}(\omega'))^\dagger\hat a_{L/Rn}^{\text{out}}(\omega)\hat a_{L/Rn}^{\text{out}}(\omega')\rangle_0. \nonumber
\end{align}

\noindent Eq. (\ref{eq:CS-corr}) can be rewritten as

\begin{equation}
    (C_{LRn}(0)+\langle J_L(0)\rangle_0\langle J_{Rn}(0)\rangle_0)^2\leq 4\langle J_{L}(0)\rangle_0^2\langle J_{Rn}(0)\rangle_0^2. \nonumber
\end{equation}

\noindent In this expression we substitute Eq. (\ref{eq:term1}-\ref{eq:term2}).
This results in

\begin{align}\label{eq:intCS}
    &\int d\omega d\omega'\langle (\hat a_L^{\text{out}})^\dagger (\hat a_{Rn}^{\text{out}})^\dagger\hat a_L^{\text{out}}\hat a_{Rn}^{\text{out}}\rangle_0^2 \\
    &-\langle (\hat a_L^{\text{out}})^\dagger(\hat a_L^{\text{out}})^\dagger\hat a_L^{\text{out}}\hat a_L^{\text{out}}\rangle_0\langle (\hat a_{Rn}^{\text{out}})^\dagger (\hat a_{Rn}^{\text{out}})^\dagger\hat a_{Rn}^{\text{out}}\hat a_{Rn}^{\text{out}}\rangle_0\leq 0. \nonumber
\end{align}

\noindent If Eq.(\ref{eq:intCS}) is violated then there exists an $\omega$ and $\omega'$ for which Eq. (\ref{eq:CSinequality}) must also be violated.
This is due to the mean value theorem for integrals.

\section{Outgoing magnon number}\label{sec:Magnonnumber}

Hawking showed in his paper \cite{Hawking1975} that the horizon of a black hole emits radiation due to the mixing of particles and antiparticles.
Also in our set-up, we find that the operators mix and in this appendix we show that the number operator in vacuum is non-zero, similar to what Hawking found for a black hole.
However unlike a black hole, the radiation spectrum is not thermally distributed.

Consider a set-up where both FMs are aligned with their respective magnetic fields or, equivalently, without driving.
This computation has been done in Appendix \ref{noSOT}.
The results of this computation aid us in contrasting the results.
Eq. (\ref{S-matrixOpnoSOT}) is the $S$-matrix for the simple set-up.
Define $|0,\text{in}\rangle$ as the vacuum-in-state that satisfies $\hat a^{\text{in}}_{\nu}|0,\text{in}\rangle=0$. 
The number of outgoing magnons on the left and right side is defined by the operator $\hat n^L(\omega)=(\hat a^{\text{out}}_{L}(\omega))^{\dagger}\hat a^{\text{out}}_{L}(\omega)$ and $\hat n^R(\omega)=(\hat a^{\text{out}}_{R}(\omega))^{\dagger}\hat a^{\text{out}}_{R}(\omega)$.
Then the expectation value of the number of outgoing magnons is

\begin{subequations}
\begin{align}
    \langle \hat n^L(\omega)\rangle_0= & 0, \\
    \langle \hat n^R(\omega)\rangle_0=& 0. 
\end{align}
\end{subequations}

\noindent As one would expect, there are no outgoing magnons when there are no ingoing magnons.
A similar computation is performed for the set-up featuring driving.
The number of outgoing magnons on the left and right FM is defined by the operator $\hat n^L_l(\omega)=(\hat a^{\text{out}}_{L}(\omega))^{\dagger}\hat a^{\text{out}}_{L}(\omega)$ and $\hat n^R_+(\omega)=(\hat a^{\text{out}}_{Rn}(\omega))^{\dagger}\hat a^{\text{out}}_{Rn}(\omega)$.

\begin{subequations}
\begin{align}
    \langle \hat n^L_l(\omega)\rangle_0\equiv & \langle0,\text{in}|(\hat a^{\text{out}}_{L}(\omega))^{\dagger}\hat a^{\text{out}}_{L}(\omega)|0,\text{in}\rangle, \nonumber\\
    = & |T(\omega)|^2,\label{eq:numberl0} \\
    \langle \hat n^R_+(\omega)\rangle_0\equiv & \langle 0,\text{in}|(\hat a^{\text{out}}_{Rn}(\omega))^{\dagger}\hat a^{\text{out}}_{Rn}(\omega)|0,\text{in}\rangle,\nonumber\\
    =& |T(\omega)|^2. \label{eq:numberr0}
\end{align}
\end{subequations}

\noindent We used Eq. (\ref{S-matrixOp}) to transform the outgoing scattering operators with ingoing scattering operators.
The number of outgoing magnons is non-zero, as is evident from the result above.
Since there are no ingoing magnons, we can only deduce that the outgoing magnons originated from the interface itself.
Also, both left and right number operators measure on average the same amount of outgoing magnons.
Thus, there is equal spontaneous emission from both sides of the interface, which implies the creation of pairs of magnons.
The spectrum is given by the transmission coefficient $|T(\omega)|^2$.
This spectrum is thermal for linearly-dispersing quasi-particles, but is not thermal in general.

\nocite{*}

\bibliography{apssamp}

\end{document}